\documentstyle[prl,aps,multicol]{revtex}
\def\beq{\begin{equation}}
\def\eeq{\end{equation}}

\def\sU{{\sf U}}
\def\1{\mbox{\small1\hskip-0.35em\normalsize1}}
\def\6{\langle }
\def\9{\rangle }

\def\tr{{\rm tr}}

\begin{document}

\title{Sufficient conditions for a disentanglement}

\author{Tal Mor\thanks{
{\sl Electrical Engineering Department, University of California, Los
Angeles, CA 90095, USA};
Electronic address: talmo@ee.ucla.edu}
and Daniel R. Terno\thanks{
{\sl Department of Physics, Technion---Israel Institute of Technology,
32\,000 Haifa, Israel};
Electronic address: terno@physics.technion.ac.il
}}


\maketitle

\begin{abstract}

We consider a disentanglement process in which 
local properties of an entangled state are preserved, while the entanglement
between the subsystems is erased. 
Sufficient conditions for a perfect disentanglement
(into product states and into separable states)
are derived, and connections to the conditions for perfect
cloning and for perfect broadcasting are observed.

\end{abstract}

\begin{multicols}{2}

Since the introduction of the paradox of Einstein, Podolsky and
Rosen~\cite{epr}, and the Schr\"{o}dinger's cat~\cite{schr}, the entanglement of
quantum states, i.e., the ability of composite quantum systems to exhibit
non-local correlations, was one of the most fascinating consequences
of the quantum formalism. Today the entanglement also serves as a basic resource
of quantum information theory~\cite{ben}, 
which is based upon
the idea that the carriers of information are physical objects and as such
are subjected to the laws of quantum mechanics. 

Being of such an importance both for the foundations of quantum mechanics 
and in  the quantum information theory, the phenomen of
entanglement  has been extensively studied.  Recent studies include  characterization and
classification of the entanglement,  manipulation of the
entanglement, and its potential applications~\cite{list}. The interplay of the local and global properties
of the composit systems is both  interesting and important in this context. One of the 
questions that can be asked is a possibility of  erasing the
entanglement between the subsystems, while keeping their local properties
intact~\cite{ter}. 

Two different kinds of the {\em disentanglement} procedure were suggested~\cite{ter}:
 disentanglement to the product state of the reduced density matrices,
\beq
\rho\rightarrow \rho_{dis}=\tr_B\rho\otimes\tr_A\rho,
\eeq
and disentanglement into  separable states,
\beq
\rho\rightarrow \rho_{dis}=\sum_i w_i\rho_i^A\otimes\rho_i^B,
\eeq
with $\rho_{dis}$ being a state that satisfies 
$\tr_A\rho=\tr_A\rho_{dis},~\tr_B\rho=\tr_B\rho_{dis}$, thus,
a state which has the same local properties as the original state $\rho$.
As it was shown by Terno~\cite{ter} and Mor~\cite{mor} an
unknown entangled state cannot be disentangled neither to  product, nor
to  separable states.

Another question is a possibility of a state-dependent disentanglement. By
this we understand a procedure that operates on a secretly chosen state, $\rho_i$,
which belongs to some predefined set of entangled states $\{\rho^1,\ldots,
\rho^n\}$. The output of this process is a corresponding state
$\rho^i_{dis}$.

Several sets were analyzed as possible inputs of a state-dependent
disentanglement machine. One of the  sets that cannot be disentangled into
product states consists of the two possible outputs of the state-dependent
cloning machine~\cite{dagmar}, which operates obliviously on two arbitrary
pure states $\psi_1$ and $\psi_2$. Their optimal copies cannot be
disentangled into product states, since it would increase their
distinguishabiltity beyond its optimal value~\cite{ter}.

The states
\begin{eqnarray}
|\psi_0 \rangle &=&
c_\phi
{c_\theta \choose s_\theta}
{c_\theta \choose s_\theta}
  +              s_\phi
{s_\theta \choose - c_\theta}
{s_\theta \choose - c_\theta}
\nonumber \\
|\psi_1 \rangle &=&
c_\phi
{c_\theta \choose - s_\theta}
{c_\theta \choose - s_\theta}
  +              s_\phi
{s_\theta \choose c_\theta}
{s_\theta \choose c_\theta}
\ ,
\end{eqnarray}
with $c_\phi \equiv \cos \phi$, etc., also cannot be disentangled into product
states~\cite{ter,mor}.

On the other hand,  pairs of maximally entangled states,
not necessary orthogonal, can be
disentangled~\cite{mor}.

The following set of states~\cite{mor}
\begin{eqnarray}
|\psi_0 \rangle &=& |00\rangle  \nonumber \\
|\psi_1 \rangle &=& |11\rangle  \nonumber \\
|\psi_2 \rangle &=& (1/\sqrt2)|00\rangle + |11\rangle
\end{eqnarray}
can be
disentangled into a mixture of tensor product states, but not into product
states, and finally,
the following set of states~\cite{mor}
\begin{eqnarray}
|\psi_0 \rangle &=& |00\rangle  \nonumber \\
|\psi_1 \rangle &=& |11\rangle  \nonumber \\
|\psi_2 \rangle &=& (1/\sqrt2)[ |00\rangle + |11\rangle ] \nonumber \\
|\psi_3 \rangle &=& (1/\sqrt{2}) |++\rangle ,
\end{eqnarray}
where $|+\9=(1/\sqrt{2}) ( |0\9+|1\9)$,
cannot be disentangled at all.

An approximate disentanglement~\cite{kar},
and disentanlement restricted to local processes~\cite{kar2} 
were recently also investigated.

In this paper we present sufficient conditions for the exact
disentanglement.
The conditions for disentanglement into product states correspond to a
sufficient conditions for cloning of one of the subsystems~\cite{bar:1}, 
and 
the condition for disentanglement into separable states corresponds to a
sufficient condition for broadcasting~\cite{bar:1} of one of the subsystems.

{\em Proposition 1}.---
\newline
{\em 1a: } Any set of perfectly
distinguishable states can be disentangled.
\newline
{\em 1b: } Any set of states
with identical reduced density matrices can be
disentangled.
\newline
{\em Proof}.---  
In the former case the
disentanglement procedure consists in an identification of the
state, followed by the bilocal preparation of the
corresponding reduced density matrices. In the later case, the
same bilocal preparation is performed for any of the possible
inputs. QED

A nice corollary of this result is that {\em any set of maximally
entangled states can be disentangled}. It is so because a reduced density
matrix of any maximally entangled state is a maximally mixed
state,~$\1/n$, where $\1$ is a unit matrix and $n$ is a
dimensionality of the subsystem.

It is interesting to note that the first part of the Proposition corresponds to
the ``if'' part of the no-cloning theorems~\cite{clon},
namely, that perfectly distinguishable states 
can be cloned, and both parts correspond to the ``if'' part of the no-cloning theorem
for mixed states~\cite{bar:1}, namely, that identical or orthogonal
mixed states can be cloned.

To derive a sufficient condition for the disentanglement into
separable states, we need to present here some properties of the
broadcasting~\cite{bar:1,cr} of a quantum state onto
two separate quantum systems. After the broadcasting the reduced density matrices
of each of the subsystems are identical with the broadcasted state, which is destroyed. This
procedure has at the input an unknown state $\rho$ from the known list and an ancilla
in some standard state $\Upsilon$. Its output is some state $\tilde{\rho}$,
$\rho\otimes\Upsilon\rightarrow\tilde{\rho}$, which satisfies
\beq
\tr_A\tilde{\rho}=\tr_B\tilde{\rho}=\rho.
\eeq
A necessary and sufficient condition for the broadcasting
of a  set is that the density matrices of its states
commute.

For a perfect disentanglement we need only the `if' part of this result. If density
matrices commute, they can be diagonalized simultaneously, therefore
we write them in their eigenbasis. The standard ancilla state is taken to
be $\Upsilon=|0\9\6 0|$.
As a result, any unitary transformation which is consistent with
\beq
|0\9|0\9\rightarrow|0\9|0\9, \ {\rm and}~ |1\9|0\9\rightarrow |1\9|1\9,
\eeq
does the job. In particular,
\beq
\sU=\sU^{\dag}=\left(
\begin{array}{cccc}
1 & 0 & 0 & 0 \\
0 & 1 & 0 & 0 \\ \label{un}
0 & 0 & 0 & 1 \\
0 & 0 & 1 & 0 \\
\end{array} \right)
\eeq
is such a transformation. Now we are able to prove the following

{\em Proposition 2}. Any set of entangled qubits can be
disentangled into separable states if the reduced density matrices
of one of the parties (Alice or Bob) commute.
\newline
{\em Proof}.---
Let us suppose that Bob's reduced density matrices commute and
choose the local basis where they are diagonal. Then
the most general form of a density matrix that can belong to
this set is
\beq
\rho^{AB}=\left(
\begin{array}{cccc}
a &  b & c & d \\
b^* & e & f & g\\
c^* & f^* & h & -b\\
d^* & g^* & -b^* &s\\
\end{array}\right), \label{fr}
\eeq
where $s=1-a-e-h$
and the matrix
is a subject to the positivity constraints.
Bob's reduced density matrix is
\beq
\rho^{B}=\left(
\begin{array}{cc}
a+h &  0  \\
0   & e+s \\
\end{array}\right). \label{br}
\eeq

We append the ancilla $C$ in the state $|0\9\6 0|$ at Bob's side and perform
a local broadcasting [with the operator $\sU$ of Eq.~(\ref{un}) with Bob's
bit first (control bit) and the
ancilla bit second (target bit)],
\begin{eqnarray}
\lefteqn{~~~~~~~~~~~~~~~~~~~~~~~~~\rho^{AB}\otimes|0\9\60| \rightarrow } \nonumber    \\
& & \tilde{\rho}^{ABC}  =  (\1^A\otimes\sU^{BC})(\rho^{AB}\otimes|0\9\60|)(\1^A\otimes\sU^{BC}).
\end{eqnarray}
Now we can trace out
either Bob's particle or the ancilla. The
result is the same and the output is
\beq
\rho_{out}=\left(
\begin{array}{cccc}
a & 0 & c & 0\\
0 & e & 0 & g\\
c^*&0 & h & 0\\
0 & g^* & 0 & s\\
\end{array}\right),
\eeq
which obviously has the same reduced density matrices as its
original from the Eq.~(\ref{fr}).

It remains to show that $\rho_{out}$ is a separable state.
For $2\times2$ density matrices a positive partial transposition is
a necessary and sufficient condition of a separability~\cite{sep}.
Namely, a density matrix $\rho$ is separable if and only if
the matrix which results from its partial transposition is positive, i.e.,
represents a valid physical state.

 We perform a partial transposition  on the second system and
get
\beq
\sigma=\rho^{T_2}_{out}=\rho_{out},
\eeq
which is identical with the original density matrix. Thus
 $\rho_{out}$ is always separable.
QED

It is easy to see why neither of the above sufficient conditions is not
a necessary one. For example, reduced density matrices of the states that
are already separable need not to commute. However, the procedure
of their `disentanglement' consists in doing nothing--- $\sU=\1$.

\bigskip\noindent{\bf Acknowledgments}\medskip

The question of necessary and sufficient conditions for the disentanglement
was first posed by Chris Fuchs, and the possible connection to the broadcast issue
was first suggested by Charles Bennett. We greatfully acknowledge
very useful discussions with 
Charles Bennett, Dagmar Bru{\ss}, Chris Fuchs, and
Lev Vaidman.

DRT was supported by a grant
from the Technion Graduate School.
The work of TM was supported 
in part by grant \#961360 from the Jet
Propulsion Lab,
and grant \#530-1415-01 from the DARPA Ultra program.

\end{multicols}
\end{document}